# Nano-sheets of two-dimensional polymers with dinuclear (arene)ruthenium nodes, synthesised at a liquid/liquid interface

Ana Cristina Gómez Herrero,[1] Michel Féron,[2] Nedjma Bendiab,[1] Martien Den Hertog,[1] Valérie Reita,[1] Roland Salut,[2] Frank Palmino,[2] Johann Coraux,*[1] and Frédéric Chérioux *[2]

[1] Univ. Grenoble Alpes, CNRS, Grenoble INP, Institut NÉEL, F-38000 Grenoble, France.
[2] Univ. Bourgogne Franche-Comté, FEMTO-ST, UFC, CNRS, 15B avenue des Montboucons, F-25030 Besançon Cedex, France.

* johann.coraux@neel.cnrs.fr ; frederic.cherioux@femto-st.fr

**Abstract**

We developed a new class of mono- or few-layered two-dimensional polymers based on dinuclear (arene)ruthenium nodes, obtained by combining the imine condensation with an interfacial chemistry process, and use a modified Langmuir-Schaefer method to transfer them onto solid surfaces. Robust nano-sheets of 2D polymers including dinuclear complexes of heavy ruthenium atoms as nodes were synthesised. These nano-sheets, whose thickness is of a few tens of nanometers, were suspended onto solid porous membranes. Then, they were thoroughly characterised with a combination of local probes, including Raman scattering, Fourier transform infrared spectroscopy and transmission electron microscopy in imaging and diffraction mode.

**Introduction**

In the past decade, nano-sheets of two-dimensional (2D) materials with thickness of few tens of nanometers have attracted much attention because of their appealing properties. These multi-layered materials are candidates for catalytic, sensing, magnetic materials, for adsorption and separation functions, or for energy applications.[1] Bottom-up approaches to their synthesis were developed rather recently. Chemical and physical vapour deposition have been mainly used for inorganic compounds, while wet chemistry has been employed to deliver easily processable and often organic (but not exclusively) nano-sheets, especially of 2D polymers.[2] This approach is very powerful owing to the versatility in the choice of the possible starting building-blocks (metal ions, molecules, etc.). Several routes have been explored: (i) at air/solid and ultrahigh vacuum/solid interfaces with the surface synthesis of covalent-organic, metal-organic or metal-ligand frameworks,[3,4] (ii) at air/liquid interfaces with or without an external stimulus to trigger the polymerisation[5,6] and (iii) at liquid/liquid interfaces.[7] The latter implements an interfacial polymerisation of two precursors, each being soluble only in one of the two immiscible liquids. The synthesis of the 2D nano-sheets is thus confined to the interface between the two liquid phases. These methods have rarely provided polymers with high crystallinity on a few-1 μm² area and with a thickness down to 2 nm. In the literature many of the reported 2D nano-sheets are organic-only polymers (i.e., free of transition metal atoms), and in some cases they include metallic centres, brought in, in the form of single transition metal atoms, by so-called mononuclear monomer building-blocks.[8] Dinuclear 2D nano-sheets, comprising dinuclear metallic complexes (i.e. metal-organic fragments each comprising two transition metal atoms), have not been reported to our knowledge. Dinuclear complexes, implemented within small and larger molecules (see, as an example, the compounds shown in Scheme 1), have however received special attention for decades as efficient catalysts.[9] Whether such functionality could be given to polymers and 2D nano-sheets in particular seems an exciting question. More generally, dinuclear species present special properties and functions unlike those of mononuclear species, which are controlled by the nature of the ligands bridging the metal atoms and of the cyclic molecule bonded to the metal atoms, and by the kind of metal atom.[10] These properties are now exploited for new generations of devices in photocatalytic water splitting, biosensors, optoelectronics etc.[10]

Here, we introduce a heavy-element, ruthenium, into the skeleton of 2D polymers. For that purpose we exploit a Schiff-base condensation between an amine as linker and bipyramidal $Ru_2S_3$ coordinated with two arene ligands as nodes. By using interfacial chemistry at a liquid/liquid interface, robust nanosheets of 2D polymers were obtained. We demonstrate the suspension of the multilayered 2D polymers across electron microscopy grids, and the transfer to solid substrates, following a modified Langmuir-Schaefer method.

## 2. Methods

### 2.1. Synthetic procedures

All reactions were carried out under nitrogen, by using standard Schlenk techniques. Solvents were degassed prior to use. The dinuclear dichloro complexes $[Ru(C_6Me_6)Cl_2]_2$ and $[Ru_2(C_6Me_6)_2(p\text{-}S\text{-}C_6H_4\text{-}Br)_3]Cl$ were synthesised by previously described methods.[11] All other reagents were purchased (Sigma Aldrich, or Strem) and used as received. The silica gel used for column chromatography was purchased from Merck. The deuterated NMR solvents were purchased from Euriso-top. The NMR spectra were recorded using a Bruker Avance-400 MHz spectrometer.

The tris(bromophenyl)dinuclear complex, $[Ru_2(C_6Me_6)_2(p\text{-}S\text{-}C_6H_4\text{-}Br)_3]Cl$ (0.1 mmol) and 4-formylphenyl boronic acid pinacol ester (0.33 mmol), were dissolved in tetrahydrofuran. Then, an aqueous solution of $Cs_2CO_3$ (1 mL), and $Pd(PPh_3)_4$ catalyst (0.01 mmol, 11 mg) was added. The resulting mixture was refluxed in dimethylformamide for 48 hours (see Scheme 1). After cooling to 20 °C, the red solution was filtered through Celite and the solvent was removed under reduced pressure. The obtained oil was purified by column chromatography (silica gel, dichloromethane/ethanol 10:1; rf close to 0.8). Then, $[Ru_2(C_6Me_6)_2(p\text{-}S\text{-}C_6H_4\text{-}C_6H_4\text{-}CHO)_3]Cl$ (1) was isolated after evaporation of the solvent, as a red-orange powder.

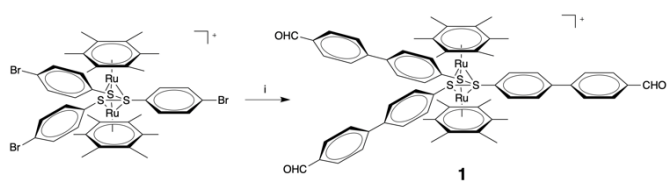

**Scheme 1** Synthesis of $[Ru_2(C_6Me_6)_2(p\text{-}S\text{-}C_6H_4\text{-}C_6H_4\text{-}CHO)_3]Cl$. i: 4-formylphenyl boronic acid pinacol ester, $CS_2CO_3$, $Pd(PPh_3)_4$, DMF, reflux, 48h.

The $[Ru_2(C_6Me_6)_2(p\text{-}S\text{-}C_6H_4\text{-}C_6H_4\text{-}CHO)_3]$ cation **1** is unambiguously characterised by its $^1H$ and $^{13}C$ NMR spectra.
$^1H$ NMR (400 MHz, $CDCl_3$, 21 °C): δ = 1.58 (s, 36H, $CH_3$-Ar), 7.81 (d, $^3J_{H;H}$ = 8.2 Hz, 6H), 7.82 (d, $^3J_{H;H}$ = 8.2 Hz, 6H), 7.96 (d d, $^3J_{H;H}$ = 8.2 Hz, 6H), 7.98 (d, $^3J_{H;H}$ = 8.2 Hz, 6H), 10.02 (s, 3H, -CHO). $^{13}C$ NMR (100 MHz, $CDCl_3$, 21 °C): δ = 15 ($CH_3$-Ar), 95 (Ru-C-Ar), 127 (C-Ar), 130 (C-Ar), 135 (C-Ar), 140 (C-Ar), 201 (C-Ar), 135.48 (C=O).

To characterise the formation of the imine bonds, the FT-IR spectrum of the polymer (compound of Fig. 1b) was compared with that of a reference compound (label 2 in Scheme 2), which presents the main chemical groups expected in the polymer. The latter compound is obtained by reacting monomer 1 with 4-cyanoaniline (Scheme 2). The trisbromodinuclear complex (1), $[Ru_2(C_6Me_6)_2(p\text{-}S\text{-}C_6H_4\text{-}C_6H_4\text{-}CHO)_3]Cl$ (0.1 mmol) and 4-cyanoaniline ($NH_2\text{-}C_6H_4\text{-}CN$; 100 mmol) were dissolved in ethanol. The resulting mixture was refluxed overnight. After cooling to 20 °C, the red solution was filtered through Celite and the solvent was removed under reduced pressure. The obtained oil was washed with diethylether (3x50 ml). The obtained red powder was purified by column chromatography (silica gel, dichloromethane/ethanol 10:1; rf close to 0.8). The $[Ru_2(C_6Me_6)_2(p\text{-}S\text{-}C_6H_4\text{-}C_6H_4\text{-}CH=N\text{-}C_6H_4\text{-}CN)_3]Cl$ compound was isolated after evaporation of the solvent as a red-orange powder.

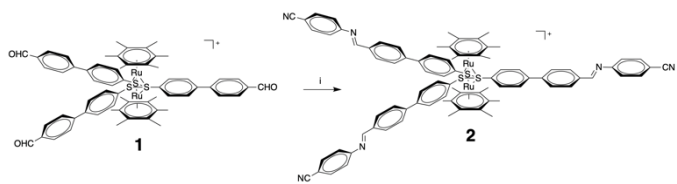

**Scheme 2** Synthesis of [Ru$_2$(C$_6$Me$_6$)$_2$(p-S-C$_6$H$_4$-C$_6$H$_4$-CH=N-C$_6$H$_4$-CN)$_3$]Cl. i: *4*-cyanoaniline, Ethanol, reflux, 2H.

The [Ru$_2$(C$_6$Me$_6$)$_2$(*p*-S-C$_6$H$_4$-C$_6$H$_4$-CH=N-C$_6$H$_4$-CN)$_3$]Cl cation **2** is unambiguously characterised by its $^1$H spectrum.
$^1$H NMR (400 MHz, DMSO-d$_6$, 21 °C): δ = 1.58 (s, 36H, CH$_3$-Ar), 7.81 (d, $^3J_{H;H}$ = 8.2 Hz, 6H), 7.82 (d, $^3J_{H;H}$ = 8.2 Hz, 6H), 7.90-8.00 (m, 18H), 8.06 (s, 3H, -CH=N-), 8.18 (d, $^3J_{H;H}$ = 8.2 Hz, 6H).

*2.2 Interfacial synthesis*

Nano-sheets of 2D polymer were synthetised at the interface of two immiscible liquids. [Ru$_2$(C$_6$Me$_6$)$_2$(p-S-C$_6$H$_4$-C$_6$H$_4$-CHO)$_3$]Cl (0.5 mg) was dissolved in a 20 mL of a mixture of ethylacetate and chloroform (18:2 mL). 1,4-benzenediamine (58 µg) was dissolved in 20 mL of water. The organic solution was deposited very carefully on the aqueous water. The duration of reaction was 1 h for the deposition of an ultra-thin film onto a TEM grid and 24 h to obtain a large quantity of 2D polymer.

*2.3. Characterisation of the samples*

Scanning electron microscopy (SEM) images were collected using a Dual Beam SEM/FIB FEI Helios Nanolab 600i system. For these measurements the polymer was deposited on Si$_3$N$_4$ TEM grids. To minimise the charge effects, a low voltage of 1 kV and a low current of 21 µA were used. The secondary electrons were detected by the TLD (Through The Lens detector) and the scanning speed optimised to reach a compromise between reduction of charge effects and preservation of a good signal-to-noise ratio. Transmission electron microscopy (TEM) and selected-area electron diffraction analysis (SAED) were carried out using a CM300 microscope equipped with a LaB$_6$ thermionic emitter operated at 100 kV.

Raman spectroscopy was performed with a 532 nm Nd:YAG laser using a confocal WITEC spectrometer at room temperature under ambient conditions. A 633 nm wavelength laser was also tested to make sure that no specific resonance of some vibration mode would enhance their signal. The signal was collected through a 50x objective with a numerical aperture of 0.75, which leads to a laser spot size of ~ 1 µm. Powders of compounds 1 and 2 were analysed, to have a reference spectrum before the reaction. Then, the polymer suspended on Si$_3$N$_4$ TEM grids was analysed as well. For the Raman spectra, the power density was varied between 0.05 and 2.0 mW/µm², to determine the onset of degradation of the samples under the laser spot and to optimise counting times. An acquisition time of 300 s was needed to measure a spectrum for powder samples and 1000 s for a suspended sample. In all Raman measurements a 1800 lines/mm grating was used in the spectrometer.

The infrared spectra were recorded with a Spectrum Two FT-IR spectrometer from Perkin Elmer.

For the atomic force microscopy (AFM), nano-sheets of 2D polymer were synthesised at the interface of two immiscible liquids as described in 2.2 during 1 h. Then, the solid polymers were mixed in 40 mL of dimethylformamide (DMF) and sonicated in a sonicator (Branson sonfier) with a fixed frequency of 20 kHz. The outpower of the sonicator was 50% of 750 W during 36 min (1080 cycles). After sonication, the mixture was centrifugated at 6000 rpm during 15 min by using a Sigma 3-30KS centrifuge. The supernatant was spin-coated on highly oriented pyrolytic graphite (HOPG) using a Laurell spin coater at a speed of 5000 rpm during 20s. Then, the sample was heated during 30 min at 80°C. The AFM measurements were then performed with a Bruker Icon AFM connected to a Nanoscope V controller. The AFM was employed in Peak-Force tapping mode using Scanasyst-Air-HPI Bruker probes of nominal stiffness of 0.4 N m-1and a nominal tip radius of 2 nm.

**3 .Results and discussion**

*3.1. Molecular design*

2D polymers containing metal atoms in their skeleton are promising candidates for physical/chemical effects and applications (spintronics, magnetism, catalysis, thermoelectricity), where the spin-orbit interaction, phonon scattering, or enhanced chemical reactivity play crucial roles.[12] Star-shaped building blocks may additionally lead to a strong enhancement of some of the polymer's physical properties, *e.g.* bringing nonlinear optical susceptibilities or electronic conductivities in hyperbranched conjugated polymers.[14] To combine this specific shape with metal atoms into a 2D polymeric sheet, a coordination reaction is often used.[13]

In the present work, an alternative strategy has been considered. Star-shaped dinuclear organometallic complexes were used as monomers. The living end-groups (available for polymerisation) of such monomers can be chosen almost at will, so in principle a broad variety of already-established robust polymerisation schemes is accessible. Here, the quantitative and versatile reaction of di-μ-chlorobis[(arene)chlororuthenium(II)] and thiophenolato derivatives has been used to first synthesise star-shaped dinuclear organometallic entities. These units are based on a closed trigonal bipyramid $Ru_2S_3$ framework which are surrounded by two arene ligands.[15] Such ligands can be designed to promote the solubility of the final molecular complex, and even of 2D polymers by decreasing electrostatic interactions between molecules or polymeric nanosheets.[16] Using the reaction between -SH and di-μ-chlorobis[(arene)chlororuthenium(II)], many types of functional group can be introduced on the thiophenolato side.

Functional group suited for a Schiff-base condensation reaction, i.e. aldehyde moieties that will react with amine groups of another monomer, were chosen here to polymerise the system via the formation of imine bonds. This reaction has recently allowed to implement dynamic covalent chemistry concepts[17,18] for the synthesis of 2D polymers.[19] The key idea is that with the chosen reaction, covalent bonds can both form and break, under equilibrium control. In our case, one monomer is hence a specifically-designed $[Ru_2(C_6Me_6)_2(p\text{-}S\text{-}(C_6H_4\text{-}C_6H_4\text{-}CHO)_3]Cl$ complex (Fig. 1a). This compound combines the $Ru_2S_3$ trigonal core and three aldehyde moieties. This complex was produced in the form of a pure salt, at the gram-scale, following a three-step-procedure with good yields (see details about the synthetic procedure in the methods section).

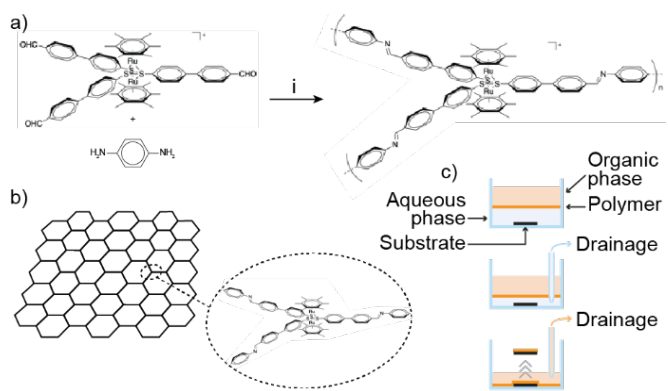

**Figure 1** a) Reaction scheme leading to a 2D Ru-imine polymer from $Ru_2(C_6Me_6)_2(p\text{-}S\text{-}(C_6H_4\text{-}C_6H_4\text{-}CHO)_3]Cl$ and 1,4-benzendiamine. i: ethylacetate/water biphasic mixture, room temperature. b) Idealized cartoon representing the 2D polymer. c) Schematics of the transfer process, inspired by the Langmuir-Schaefer method.

*3.2. Molecular reaction*

Our $[Ru_2(C_6Me_6)_2(p\text{-}S\text{-}(C_6H_4\text{-}C_6H_4\text{-}CHO)_3]Cl$ monomer was reacted with a second monomer, 1,4-benzendiamine, to form the imine bonds via a triple Schiff-base condensation. The same condensation reaction was used to form our reference compound (not a polymer; used for comparison purposes to characterise the formation of imine bonds), $[Ru_2(C_6Me_6)_2(p\text{-}S\text{-}C_6H_4\text{-}C_6H_4\text{-}CH=N\text{-}C_6H_4\text{-}CN)_3]Cl$, which was obtained by reaction of our dinuclear monomer with 4-cyanoaniline (see methods section). This dinuclear complex was obtained in the form of an orange powder, well-soluble in alcohols.

*3.3 Interfacial reaction*

The synthesis of the polymeric nano-sheets was performed at the liquid/liquid interface between an organic top phase (lower density), containing the $[Ru_2(C_6Me_6)_2(p\text{-}S\text{-}(C_6H_4\text{-}C_6H_4\text{-}CHO)_3]Cl$ complex in a chloroform/ethyl acetate mixture, and a bottom aqueous phase containing, the 1,4-benzenediamine (Fig. 1a-c, see details in the methods section).

After reaction at room temperature, a thin orange film covering an area of few cm², is visible at the interface of the two immiscible liquids. This solid thin film can be fished or transferred onto a substrate. For the latter, a variant of the Langmuir-Schaefer method (Fig. 1c) was used. Before introducing the two liquid phases, the host (clean) substrate was placed at the bottom of the beaker. Next, the interfacial synthesis was implemented as described previously. After reaction, the aqueous phase was drained to gently deposit the thin-solid orange film onto the surface of the substrate. Finally, the organic phase was completely drained too. This left the substrate covered with the polymer accessible for manipulations and characterisation.

We first address the case of a long reaction time, of 22 h, which yields a rather thick polymeric film (thickness around 150 nm). This film was fished and characterised by FT-IR spectroscopy. A clear signature of the imine bond was observed as a band centred at 1677 cm$^{-1}$ in the spectrum, and no aldehyde signature was detected (Fig. 2). This establish that the reaction sketched in Fig. 1a was successful and complete (within the detection sensitivity of our FT-IR analysis).

*3.4. Characterisation of suspended polymeric nano-sheets and the 2D polymer*

After a reaction time of 16 h, a nano-sheet of 2D polymer was obtained, which was then transferred onto a Si$_3$N$_4$ TEM grid following the above-discussed process. The process is compatible with the suspension of the polymeric nano-sheets across holes of typically 500 nm in diameter. For larger holes the nano-sheets tend to break, which might be due to surface tension issues during the transfer process.

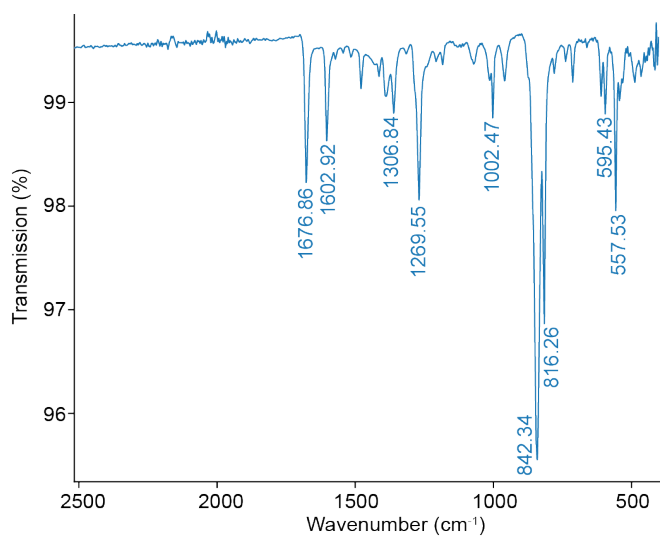

**Figure 2** FT-IR spectrum of a thin (~ 150 nm) solid film of 2D polymer. The band at 1677 cm$^{-1}$ is attributed to imine functions. No band characteristic of an aldehyde function is detected.

The SEM and TEM images (Fig. 3a-e) show the formation of solid nano-sheets covering the holes of the grids. The thickness of these suspended nano-sheets was below 70 nm. The SAED patterns consist of two rings with uniform intensity (Fig. 3f). This shows that at the scale of the electron beam (few hundred nanometers), the sheet is composed of randomly in-plane-oriented grains, or in other words, there is strong structural disorder within the electron spot size on the sample. The integrated radial scattered intensity distribution extracted from the SAED pattern (inset of Fig. 3f) exhibits a peak centred at about 11 nm$^{-1}$, corresponding to a real space distance of 0.114 nm. This distance is precisely the one measured for a lacey carbon TEM grid, and can hence be ascribed to the numerous C-C bonds in the polymer. Such signature of in-plane disorder looks similar to the one observed recently in another, Schiff-reaction-derived polymeric nano sheet.[20] In contrast, a collection of diffraction spots distributed around rings was observed in yet another Schiff-reaction-derived 2D polymer,[19] suggesting in this case that a finite number of crystalline 2D domains with random in-plane orientation contributed to the SAED pattern. Whether in our case the high degree of in-plane disorder is intrinsic to each individual polymeric layer or to the relative disorientation between successive layers cannot be determined.

To determine the thickness of an individual layer of the polymeric nano-sheet, a short-time reaction (60 min) was performed, which expectedly yields a sub-monolayer of the polymer. While such low amounts of matter could not be suspended across our TEM grids, they are readily dispersed in a DMF solvent. After spin-coating onto highly-oriented pyrolytic graphite (HOPG) and subsequent solvent evaporation, AFM reveals that the polymer forms a very flat (partial) layer on this substrate (Fig. 3g) with a characteristic phase signal (Fig. 3h). The layer's height is uniform, about 1.2 nm with respect to the substrate surface. According to this estimate, we infer that the above-mentioned 70 nm thickness for a nano-sheet corresponds to about 60 individual layers.

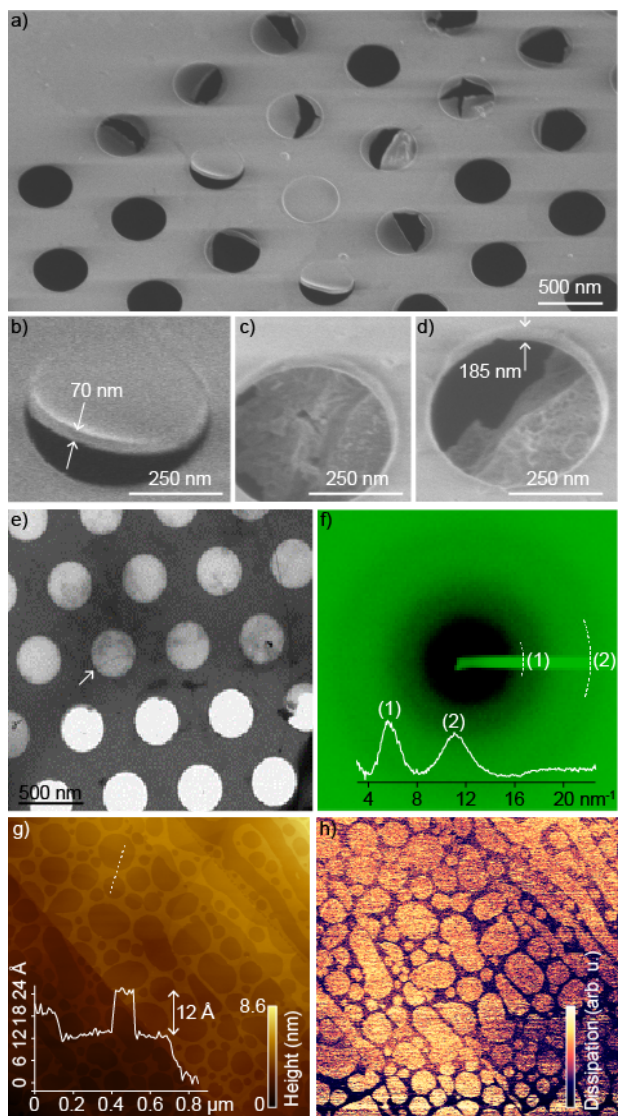

**Figure 3** a-d) Oblique-view SEM images of a polymeric nano-sheet partially suspended across a $Si_3N_4$ grid. a,b) are views of the polymer deposited atop the grid with the modified Langmuir-Shaefer method, while c,d) are views taken from below. Nano-sheets of different thicknesses, smaller than the grid's thickness, are observed. e) TEM image of a suspended polymeric nano-sheet. f) SAED pattern and its integrated radial density profile (inset), acquired on one of the nano-sheet-covered holes of the grid. The holes appearing the darkest (respectively brightest) in the SEM (respectively TEM) images are not covered by the nano-sheet. g) AFM topograph of a HOPG surface covered with a sub-monolayer of the polymer. The HOPG mono/bi/etc-atomic steps are discerned in the background of the image. The inset shows a height profile along the dotted line. (h) Dissipation signal measured while the AFM topograph was acquired, i.e. phase of the AFM signal.

The nature of the chemical bounds in the polymer was analyzed, with optical spectroscopies. FT-IR spectroscopy did not provide sufficient spatial resolution to characterise the fraction of the polymer suspended across the TEM grids' holes of the nano-sheets. We hence turned to a confocal Raman spectroscopy analysis. Let us first note that in all our measurements, and despite our efforts to minimise it by using the narrowest possible laser beam, a spurious signal is systematically produced by the $Si_3N_4$ grid, whatever the incident light power (Fig. 4). This signal is generated by the "tails" of the laser spot (focused at the optical diffraction limit), which illuminate the thick $Si_3N_4$ membrane that thereby produces the background signal. This signal is relatively strong, compared to the one produced by the suspended polymer, due to the low amount of matter at the origin of the relevant Raman scattering signal. Remember that membrane holes with diameter beyond 500 nm diameter do not allow for suspending the polymeric sheet.

In comparison, the Raman spectra acquired for powders of the precursor, $[Ru_2(C_6Me_6)_2(p\text{-}S\text{-}C_6H_4\text{-}C_6H_4\text{-}CHO)_3]Cl$, and of $[Ru_2(C_6Me_6)_2(p\text{-}S\text{-}C_6H_4\text{-}C_6H_4\text{-}CH=N\text{-}C_6H_4\text{-}CN)_3]Cl$, exhibited low background level owing to the large amount of probed matter. Note that the latter compound features the same trigonal shape, including the imine function, that should form in the polymeric nano-sheets too. Therefore, similar Raman fingerprint are expected for it and the polymer.

For the suspended nano-sheets, the Raman spectra exhibit a number of bands above background level, at 1024 cm$^{-1}$, 1081 cm$^{-1}$, 1200 cm$^{-1}$, 1290 cm$^{-1}$, 1590 cm$^{-1}$ and 1608 cm$^{-1}$. First, the spectra were compared to those of the precursor [Ru$_2$(C$_6$Me$_6$)$_2$(*p*-S-C$_6$H$_4$-C$_6$H$_4$-CHO)$_3$]Cl and of [Ru$_2$(C$_6$Me$_6$)$_2$(*p*-S-C$_6$H$_4$-C$_6$H$_4$-CH=N-C$_6$H$_4$-CN)$_3$]Cl.

Obviously, all three spectra show strong similarities. However, there are slight key differences, which are apparent within the three shaded (yellow, pink, green) rectangles in Fig. 4. In the yellow-shaded area corresponding to C-C stretching modes, the precursor has a characteristic signature in the form of a doublet of low-intensity bands (1172 cm$^{-1}$ and 1193 cm$^{-1}$). For both, [Ru$_2$(C$_6$Me$_6$)$_2$(*p*-S-C$_6$H$_4$-C$_6$H$_4$-CH=N-C$_6$H$_4$-CN)$_3$]Cl and the nano-sheets, this doublet is absent. Instead, an asymmetric band is detected, centred at a higher wavenumber (1202 cm$^{-1}$ and 1200 cm$^{-1}$ respectively). A similar observation was made in the pink-shaded area, also corresponding to C-C stretching modes: there again, [Ru$_2$(C$_6$Me$_6$)$_2$(*p*-S-C$_6$H$_4$-C$_6$H$_4$-CH=N-C$_6$H$_4$-CN)$_3$]Cl and the nano-sheets have similar kinds of signatures, contrasting with that of the monomer.

In the green-shaded area, the precursor exhibits a characteristic low-energy band at 1679 cm$^{-1}$, typically assigned to a C=O stretching mode.[16,21,22] As expected, this band is absent for [Ru$_2$(C$_6$Me$_6$)$_2$(*p*-S-C$_6$H$_4$-C$_6$H$_4$-CH=N-C$_6$H$_4$-CN)$_3$]Cl. Instead, a band at 1696 cm$^{-1}$, characteristic of a C=N stretching mode,[16,21,22] is observed. Neither the C=O nor the C=N stretching modes show up in the wavenumber range in the nano-sheet spectra.

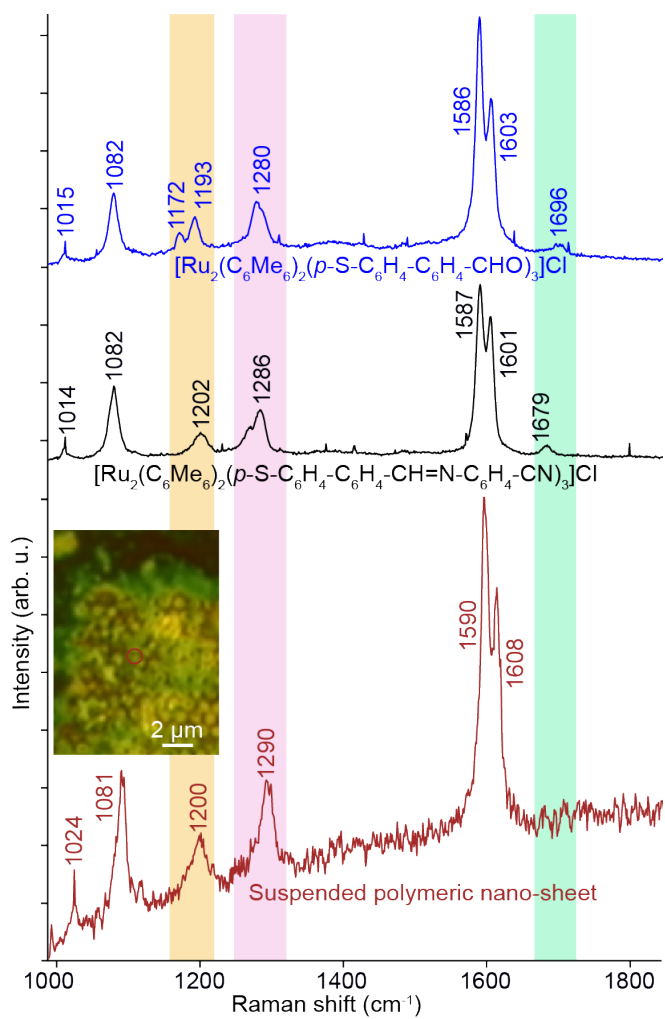

**Figure 4** Raman spectra acquired for the monomer [Ru$_2$(C$_6$Me$_6$)$_2$(*p*-S-C$_6$H$_4$-C$_6$H$_4$-CHO)$_3$]Cl (powder form), the polymer's repeat unit [Ru$_2$(C$_6$Me$_6$)$_2$(*p*-S-C$_6$H$_4$-C$_6$H$_4$-CH=N-C$_6$H$_4$-CN)$_3$]Cl (powder form) and the polymeric nano-sheet suspended across the Si$_3$N$_4$ grid, taken with a laser wavelength of 532 nm laser. The power density was 1.7 mW/mm$^2$ and 0.17 mW/mm$^2$ for the powders and polymer respectively. Acquisition times were 300 s and 1000 s for the powders and polymer respectively. The location of the laser spot, where the polymer Raman spectra was acquired, is marked with a red circle in the optical micrograph shown as the inset.

This is not a proof that the corresponding bonds are absent in the polymer, but could simply be due to a too-low signal-to-noise ratio (the bands at 1679 cm$^{-1}$ and 1696 cm$^{-1}$ in the other compounds have very low intensity). Clear signatures of another vibration mode predicted for the C=N stretching, around 1550 cm$^{-1}$,[22] were not observed either, in none of the compounds. Presumably, this is because their corresponding bands are much less intense than those for the C=C stretching, which are centred at (only) slightly higher wavenumbers (1586-1590 cm$^{-1}$, 1601-1608 cm$^{-1}$).

Summarising the information deduced from Raman spectroscopy, C-C stretching modes with a line shape and a wavenumber differing from those encountered with the monomer were observed in the polymeric nano-sheets. Interestingly, they have strong similarities with those found for the isolated repeat unit of the polymer. A specific C-C stretching mode due to the bonding with N atoms, in the -C-C=N- segments of the repeat unit and polymer can indeed be reasonably expected. This is an (indirect) indication of the formation of the C=N bridges, *i.e.* of the polymerization, following the Schiff base condensation scheme.

## 4. Conclusion

A method aiming at integrating ruthenium atoms into the skeleton of the 2D polymers was successfully developed. By combining interfacial chemistry at a liquid/liquid interface and the Schiff-base condensation, robust nanosheets of 2D polymers with bipyramidal $Ru_2S_3$ coordinated with two arene ligands as nodes were obtained. Such polymers featuring dinuclear metallic complexes could be relevant for functions in (photo)catalysis and bio sensing, in this sense inheriting the properties known for smaller kinds of molecules, but this time in a nano-sheet geometry that presents a number of practical advantages for applications. In the future, further pushing morphology and structure control down to the limit of a highly crystalline plain single-2D-sheet, for instance via an extended exploration of synthesis conditions (temperature, precursor concentrations, precursor design) or the addition of chemical substances helping the ordering of monomers at the interface could give access to much sought-after effects, e.g. band-like electronic conduction under the influence of spin-orbit coupling phenomena.

## Acknowledgments

This work was supported by the French National Research Agency through contract ORGANI'SO (ANR-15-CE09-0017). This work was partly supported by the French RENATECH network and its FEMTO-ST technological facility.

## References

(1) Ding SY, Wang W 2013 *Chem. Soc. Rev.* **42** 548; Wang J, Xu Y, Pang H 2020 *Chem. Eur. J.* **26** 6402; Zhan X, Chong Z, Zhang Q 2017 *J. Mater. Chem. A* **5** 14463; Udayabhaskararao T, Altantzis T, Houben L, Coronado-Puchau M, Langer J, Popovitz-Biro R, Liz-Marzan LM, Vukovic L, Kral P, Bals S, Klajn R 2017 *Science* 358 514; Baluri PK, Harris Samuel DG, Guruvishnu T, Aditya DB, Mahadevan SM, Udayabhaskararao T 2018 *Mater. Res. Express* **5** 014013.

(2) Dong R, Zhang T, Feng X 2018 *Chem. Rev.* **118** 6189; J. Sakamoto J, van Heijst J, Lukin O, Schluter AD 2009 *Angew. Chem. Int. Ed.* **48** 1030; Perepichka DF, Rosei F 2009 *Science* **323** 216.

(3) C. Moreno C, Vilas-Varela M, Kretz B, Garcia-Lekue A, Costache MV, Paradinas M, Panighel M, Ceballos G, Valenzuela SO, Pena D, Mugarza A 2018 *Science* **360** 100.

(4) Kambe T, Sakomoto R, Hoshiko K, Takada K, Miyachi M, Ryu JH, Sasaki S, Kim J, Nakasato K, Takata M, Nishihara H 2013 *J. Am. Chem. Soc.* **135** 2462.

(5) Liu K, Qi H, Dong R, Shivhare R, Addicoat M, Zhang T, Sahabudeen H, Heine T, Mannsfeld S, Kaiser U, Zheng Z, Feng X 2019 *Nature Chem.* **11** 994.

(6) Muller V, Hinaut A, Moradi M, Baljozovic M, Jung TA, Shahgaldian P, Mohwald H, Hofer G, Kroger M, King BT, Meyer E, Glatzel T, Schluter AD 2018 *Angew. Chem. Int. Ed.* **57** 10584.

(7) Dey K, Pal M, Rout KC, Kunjattu HS, Das A, Mukherjee R, Kharul UK, Banerjee R 2019 *J. Am. Chem. Soc.* **139** 13083; Rodenas T, Luz I, Prieto G, Seoane B, Miro H, Corma A, Kapteijn F, Llabrés i Xamena FX, Gascon J 2015 *Nature Mater.* **14** 48; Sakamoto R, Hoshiko K, Liu Q, Yagi T, Nagayama T, Kusaka S, Tsuchiya M, Kitagawa Y, Wong WY, Nishihara HA 2015 *Nat. Commun.* **6** 6713.

(8) Tran M, Kline K, Qin Y, Shen Y, Green MD, Tongay S 2019 *Appl. Phys. Rev.* **6** 041311.

(9) Süss-Fink G, Therrien B 2007 *Organometallics* **26** 766; Dyson PJ 2004 *Coord. Chem. Rev.* **248** 2443.

(10) Li G, Zhu D, Wang X, Su Z, Bryce MR 2020 *Chem. Soc. Rev* **49** 765; Shafikov MZ, Daniels R, Pander P, Dias FB, Gareth Williams JA, Kozhevnikov VN 2019 *ACS Appl. Mater. Interfaces* **11** 8182.

(11) Zelonka RA, Baird MC 1972 *Can. J. Chem.* **50** 3063; Wang JW, Moseley K, Maitlis PM 1969 *J. Am. Chem. Soc.* **91** 5970; Bennett MA, Smith AK 1974 *J. Chem. Soc., Dalton Trans.* 233.

(12) Muchler L, Casper F, Yan B, Chdov S, Felser C 2013 *Phys. Status Solidi RRL* **7** 91; Wang Y, Zhu X, Li Y 2019 *J. Phys. Chem. Lett.* **10** 4663; Afzal AM, Min KH, Ko BM, Eom J 2019 *RSC Advances* **9** 31797; Kambe

T, Sakamoto R, Kusamoto T, Pal T, Fukui N, Hoshiko K, Shimojima T, Wang Z, Hirahara T, Ishizaka K, Hasegawa S, Liu F, Nishihara H 2104 *J. Am. Chem. Soc.* **136** 14357; Wang Z, Su S, Liu F 2013 *Nano Lett.*, **13** 2842.

(13) Walch H, Dienstmaier J, Eder G, Gutzler R, Schlögl S, Sirtl T, Das K, Schmittel M, Lackinger M 2011 *J. Am. Chem. Soc.* **133** 7909; Gutzler R, Cardenas L, Lipton-Duffin J, El Garah M, Dinca LE, Szakacs CE, Fu C, Gallagher M, Vondráček M, Rybachuk M, Perepichka DFF, Rosei F 2014 *Nanoscale* **6** 2660; Sheberla D, Sun L, Blood-Forsythe MA, Er S, Wade CR, Brozek CK, Aspuru-Guzik A, Dincă M 2014 *J. Am. Chem. Soc.* **136**, 8859; Bebensee F, Svane K, Bombis C, Masini F, Klyatskaya S, Besenbacher F, Mario R, Hammer B, Linderoth TR 2014 *New J. Phys.* **53** 12955; Kumar A, Banerjee K, Foster AS, Liljeroth P 2018 *Nano Lett.* **18** 5596; Huang X, Sheng P, Tu Z, Zhang F, Wang J, Geng H, Zou Y, Di CA, Yi Y, Sun Y, Xu W, Zhu D 2015 *Nat. Comm.* **6** 7408.

(14) Dhenault C, Ledoux I, Samuel IDW, Bourgault M, Lebozec H 1995 *Nature* **374** 339; Chérioux F, Guyard L 2001 *Adv. Func. Mater.* **11** 305.

(15) Chérioux F, Therrien B, Süss-Fink G 2003 *Eur. J. Inorg. Chem.* **6** 1043; Chérioux F, Therrien B, Süss-Fink G 2004 *Inorg. Chim. Acta* **357** 834.

(16) Coraux J, Hourani W, Muller VL, Lamare S, Kamaruddin DA, L. Magaud L, Bendiab N, Den Hertog M, Leynaud O, Palmino F, Salut R, Chérioux F 2017 *Chem. Eur. J.* **23** 10969.

(17) Rowan SJ, Cantrill SJ, Cousins GRL, Sanders JKM, Stoddart JF 2002 *Angew. Chem. Int. Ed.* **41** 898.

(18) Yu Y, Bin J, Lei S 2017 *RSC Advances* **7** 11496; Ciesielski A, El Garah M, Haar S, Kovaricek P, Lehn JM, Samori P 2014 *Nature Chem.* **6** 1017.

(19) Sahabudeen H, Qi H, Glatz BA, Tranca D, Dong R, Hou Y, Zhang T, Kuttner C, Lehnert T, Seifert G, Kaiser U, Zhikun-Zheng AF, Feng X, 2016 *Nat. Commun.* **7** 13461.

(20) Liu J, Yang F, Cao L, Li B, Yuan K, Lei S, Hu W, 2019 *Adv. Mater.* **31**, 1902264.

(21) Lin-Vien D, Colthup N, Fateley W, Grasselli J. *The handbook of infrared and Raman characteristic frequencies of organic molecules*. Elsevier, 1991; B. Schrader. *Infrared and Raman spectroscopy: methods and applications*. John Wiley & Sons, 2008.

(22) Dai W, Shao F, Szczerbiński J, McCaffrey R, Zenobi R, Jin Y, Schlüter AD, Zhang W 2016 *Angew. Chem. Int. Ed.* **55** 213.